\documentclass[%
 reprint,
 superscriptaddress,
 amsmath,amssymb,
 prd,
]{revtex4-2}

\usepackage{graphicx}
\usepackage{subcaption}
\usepackage{multirow}

\usepackage{hyperref}

\usepackage{lipsum}

\begin{document}

\title{Context-Enriched Identification of Particles with a Convolutional Network for Neutrino Events}

\author{F.~Psihas}
\affiliation{Department of Physics, University of Texas at Austin\protect\\ 
Austin, Texas 78712, USA}
\author{E.~Niner}
\affiliation{Fermi National Accelerator Laboratory\protect\\ 
Batavia, Illinois 60510, USA}
\author{M.~Groh}
\affiliation{Department of Physics, Indiana University\protect\\
Bloomington, Indiana 47405, USA}
\author{R.~Murphy}
\affiliation{Department of Physics, Indiana University\protect\\
Bloomington, Indiana 47405, USA}
\author{A.~Aurisano}
\affiliation{Department of Physics, University of Cincinnati\protect\\
Cincinnati, Ohio 45221, USA}
\author{A.~Himmel}
\affiliation{Fermi National Accelerator Laboratory\protect\\ 
Batavia, Illinois 60510, USA}
\author{K.~Lang}
\affiliation{Department of Physics, University of Texas at Austin\protect\\ 
Austin, Texas 78712, USA}
\author{M.~D.~Messier}
\affiliation{Department of Physics, Indiana University\protect\\
Bloomington, Indiana 47405, USA}
\author{A.~Radovic}
\affiliation{Department of Physics, College of William \& Mary\protect\\
Williamsburg, Virginia 23187, USA}
\author{A.~Sousa}
\affiliation{Department of Physics, University of Cincinnati\protect\\
Cincinnati, Ohio 45221, USA}

\begin{abstract}Particle detectors record the interactions of subatomic particles and their passage through matter.
The identification of these particles is necessary for in-depth physics analysis.
While particles can be identified by their individual behavior as they travel through matter, the full context of the interaction in which they are produced can aid the classification task substantially.
We have developed the first convolutional neural network for particle identification which uses context information. 
This is also the first implementation of a four-tower siamese-type architecture both for separation of independent inputs and inclusion of context information.
The network classifies clusters of energy deposits from the NOvA neutrino detectors as electrons, muons, photons, pions, and protons with an overall efficiency and purity of 83.3\% and 83.5\%, respectively.
We show that providing the network with context information improves performance by comparing our results with a network trained without context information.
\end{abstract}

\maketitle

\section{Introduction}

Experiments in high-energy physics employ a variety of detection technologies and analysis techniques to study the properties of subatomic particles.
Such techniques are vital for interpreting raw data as particle interactions from which physical quantities can be extracted.
The performance of reconstruction algorithms largely dictates the physics potential of particle physics experiments.
In the past several decades, reconstruction techniques have evolved to adapt to the growing complexity of detector technologies and the increasing volume of data, from the hand-scanning of bubble chamber plates~\cite{bubbles} to the application of deep learning algorithms on large datasets~\cite{Nature}.

Classification of interactions, or events, in detector data is often similar
to image recognition in that topological features are useful discriminators for both tasks.
Detector data are very image-like in that, in many cases, they can be mapped into two-dimensional arrays of values preserving spatial information~\cite{kagan}.
In recent years, deep neural networks (predominantly convolutional, adversarial, and recurrent neural networks) have been employed in experiments across the field for tasks such as classification~\cite{eventCVN}, simulation~\cite{GANs}, and energy reconstruction~\cite{JMEnergy}, among others~\cite{Nature}.
Convolutional Neural Networks (CNNs) have proven useful for detector data analysis given their ability to exploit topological information, and are increasingly common in the field. 
In neutrino experiments, CNNs are most commonly used for classification tasks~\cite{eventCVN,microboone,next,MinervaVTX}, though other applications, including image segmentation, have been tested on simulations~\cite{microbooneSS}.


In experiments measuring neutrino oscillations, the observables of interest are the flavor 
and the energy of incident neutrinos, e.g. \cite{nova2018}, among other quantities.
Identification of the neutrino flavor requires discrimination between charged or neutral current interactions and separation from backgrounds, which include muons, electrons, photons, and hadrons from cosmogenic sources. 
Being neutral particles, neutrinos themselves deposit no energy as they traverse the detectors.
Instead, the neutrino flavor can be determined by identifying the leptonic charged particle produced in the neutrino interaction.
In addition, identifying all the final-state particles allows for cross section measurements of specific interaction types, a topic of great interest in the field~\cite{nuintreview}, and enables in-depth studies of the neutrino events, leading to improvements in energy reconstruction.


Much like in image recognition, the context within which an object is depicted can aid the classification task.
In particle identification, the context is the remainder of the event activity which corresponds to all particles produced in the interaction.
This context provides information about the physics of the interaction, such as conservation of energy and momentum, conservation of quantum numbers, nuclear effects, particle multiplicity, etc.
By providing context to the network, it is encouraged to learn from physical quantities implying conservation laws as well as the relative topology of the particle with respect to the accompanying activity.

We present the first implementation of a CNN for particle identification in neutrino detector data which uses context information.
This is also the first implementation of a CNN for pattern recognition which uses a four-tower siamese-type architecture~\cite{siamese} for including context information and independent two-dimensional views.

We have developed a CNN for classification of individual final state particles originating from neutrino interactions. 
We evaluate the performance of the technique on simulations from the NOvA experiment, but this technique is extendable to similar detector technologies.
The 
classification of particles in NOvA events and our choices for a network architecture are described in Section~\ref{sec:particles}.
We discuss the performance and utilization of this network in Section~\ref{sec:performance}.

\begin{figure*}
    \centering
    \includegraphics[width=0.99\linewidth]{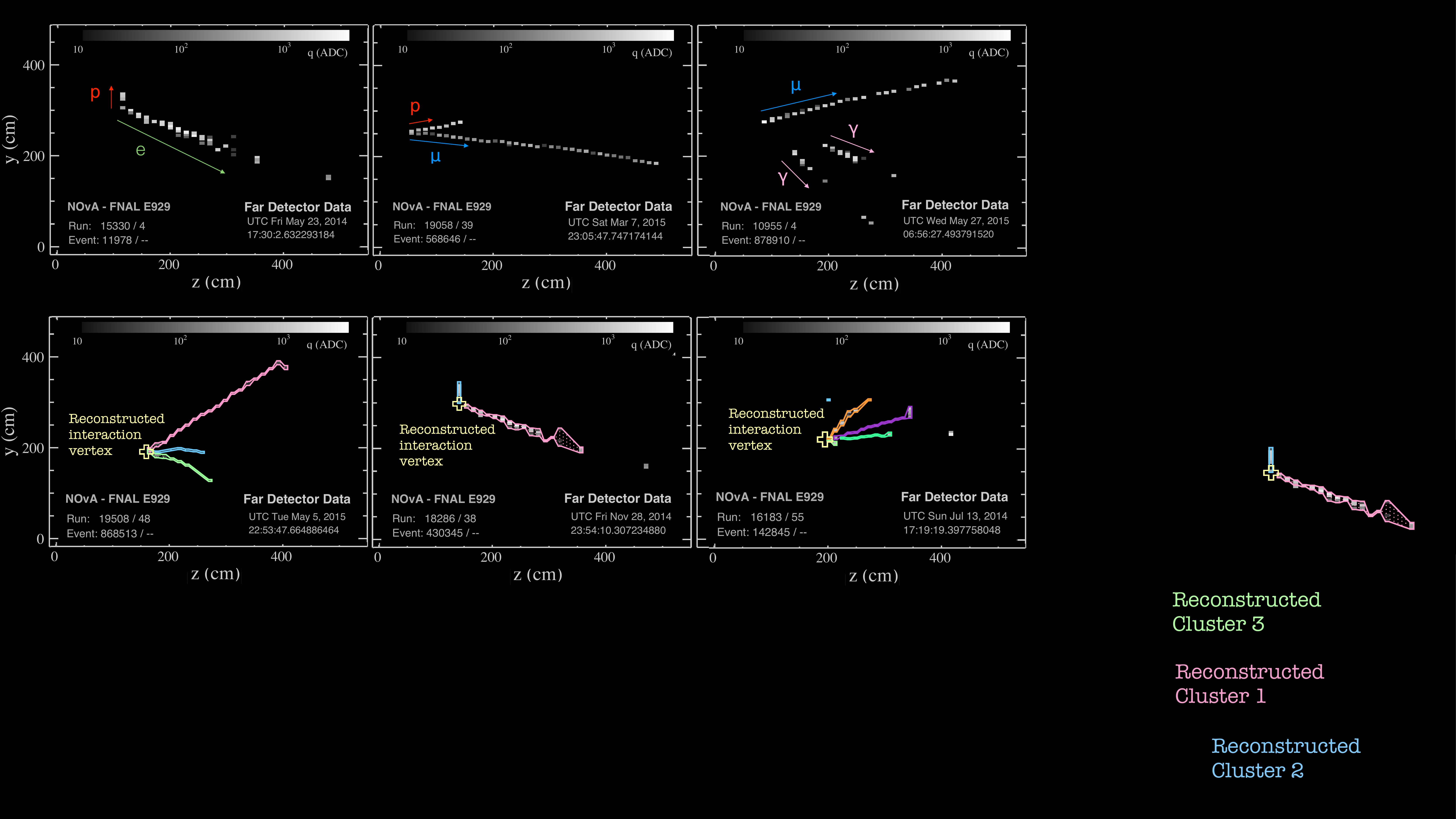}
    \caption{A Side view of six neutrino events in the NOvA detectors.
    These candidate data events were selected in NOvA's analyses.
    Top left: a $\nu_e$ interacts producing an electron and a proton.
    Top middle and right: $\nu_\mu$ interactions producing a muon accompanied by a proton or two photons from a $\pi^0$ decay, respectively.
    Bottom: From left to right, a $\nu_\mu$ event, a $\nu_e$ event and a neutral-current interaction.
    The reconstructed clusters are shown as colored regions surrounding the hits.
    }
    \label{fig:topo02}
\end{figure*}

\section{Particle Classification}
\label{sec:particles}

The NuMI Off-Axis Electron Neutrino Appearance (NOvA) experiment is a long-baseline neutrino oscillation experiment.
NOvA's primary physics goal is to measure neutrino oscillations by detecting neutrino interactions in two detectors, a near detector at 1~km and a far detector at 810~km from the muon neutrino ($\nu_\mu$) beam source~\cite{numi}.
In the far detector, a fraction of the $\nu_{\mu}$s will have oscillated into electron neutrinos ($\nu_e$) or tau neutrinos ($\nu_\tau$) with a probability dependent on the energy of the neutrino and other physical parameters associated with this phenomenon~\cite{PMNS}.
Making precise measurements of neutrino oscillation probabilities requires both identification of the flavor and measurement of the incoming neutrino energy, which in turn rely on the ability to study particle interactions in matter.


The NOvA detectors are sampling calorimeters whose detection mechanism consists of collecting the scintillation light produced by charged particles within them. The detectors are made of alternating vertical and horizontal planes filled with scintillator material~\cite{novadet} such that a view of the energy deposits, or hits, from the vertical or horizontal planes is equivalent to a two-dimensional top or side view of the activity in the detector.

Figure~\ref{fig:topo02} shows hits from particles produced in neutrino interactions in the NOvA far detector.
The energy deposited by charged particles along their path can be used to reconstruct their trajectory, or track.
The amount of energy deposited per unit distance ($dE/dx$) along tracks is an example of a useful discriminator for particle identification given the well-known behavior of particles as they travel through matter~\cite{PDGpassage}.
The events depicted are particles produced at NOvA's energy range, with a peak neutrino energy of 2~GeV.
Different particles, with characteristically varying energy depositions, may be produced in neutrino interactions.
At these energies, the signature of a muon is a long straight trajectory with a uniform $dE/dx$ profile, corresponding to the minimum ionizing energy of the particle.
Proton tracks tend to be significantly shorter than muon tracks, and are characterized by large energy deposits near the end of the track.
While proton, charged pion, and muon trajectories can look similar, the scattering behavior of each can be substantially different, causing deviations in their trajectories that are well described~\cite{PDGpassage}.
Electron and photon signatures are characterized by electromagnetic showers which broaden along the initial portions of their path.

Visual discrimination of different particles is often enhanced by contextual information from the whole interaction.
In the case of photons and electrons, both initiate electromagnetic showers which appear similar in these detectors.
However, they can be distinguished by the separation of the shower at the interaction vertex, the point where the neutrino interaction occurred.
Electrons will deposit energy immediately from the vertex while photons will travel some distance, characterized by the radiation length of the medium, before producing an electron-positron pair and initiating a shower.



\begin{figure*}
    \centering
    \includegraphics[width=1\linewidth,trim={65 200 65 0}]{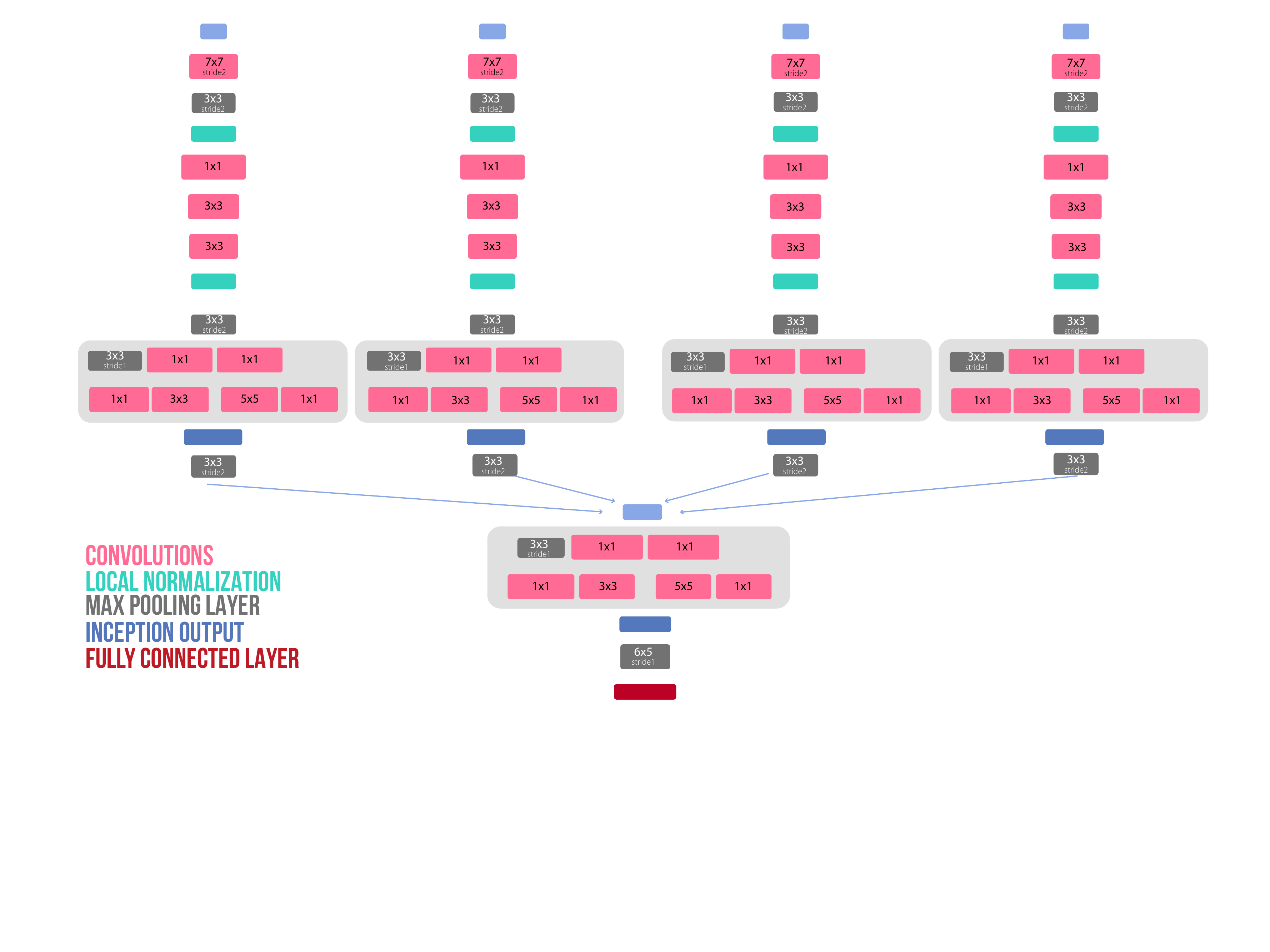}
    \caption{A diagram of the CNN architecture for particle classification.
    The network has four towers for the top and side views of the event.
    Two of the four towers are for the views of the particle of interest.
    The other two towers are for the views of the entire interaction.
    Highlighted in light gray are inception modules.}
    \label{fig:net}
\end{figure*}

Our network classifies particles isolated by the NOvA reconstruction algorithms~\cite{novareco}, which aim to separate the contributions from each individual particle into a cluster.
Figure~\ref{fig:topo02} (bottom) shows the reconstructed clusters in NOvA events.
Often, particle trajectories overlap in the same detector cell which causes individual clusters to include hits produced by multiple particles.
The average purity of the clusters varies by particle type, being typically much lower for short clusters, closest to the interaction vertex, which correspond largely to pions and protons.
Despite the inefficiencies present in the clustering, a sample of particles built from fully simulated neutrino interactions was used as a training sample as they bear a greater resemblance to detector data than what could be accomplished with simulated single particles, where the clusters would contain no contamination.
For particle classification, we employ the siamese tower structure to separate the top and side views of the cluster~\cite{eventCVN}, and we
incorporate two additional towers, one for each view of the complete interaction, thus including the full event information as context.
Several characteristics of the clusters in the event, like opening angle and distance from the interaction vertex, are contextual indicators of their identity and have been used successfully in the past as part of event selection.
The intention of adding full context images to the input is to provide the network with this capability as well.

The context of an image has been shown to be critical for object detection tasks~\cite{background} and context-enriched learning has been previously implemented for image recognition~\cite{imagecontext}.
The four-tower architecture implemented in this work is a new approach at context-enriched classification, and the first instance of context-enriched deep learning algorithms for particle physics.


\subsection{Training Methodology}
\label{sec:methods}

We use a sample of simulated neutrino interactions in the NOvA Far Detector for training, testing, and evaluation.
While raw data from the NOvA experiment is not publicly available, details of NOvA's simulation produced with GENIE~\cite{genie} and GEANT4~\cite{geant} can be found in reference~\cite{novasim}.

Hits within the interactions 
are clustered within regions of space spanning some angle with respect to the reconstructed interaction vertex using a fuzzy K-means algorithm~\cite{fuzzyk}, and matched between the top and side views by comparing energy depositions along the z-axis using a Kuiper test~\cite{ninerthesis}.
The construction of clusters is constrained by the spatial and temporal resolution of the detector and subject to inefficiencies in the algorithm, which impacts the ability to contain the detected energy deposited by a single particle uniquely.
Inefficiencies in vertex finding and separating overlapping particles
also contribute to reducing the quality, the completeness and purity, of the reconstructed clusters. 

\begin{figure*}
     \centering
     \includegraphics[width=0.49\linewidth,trim={270 75 15 0},clip=true]{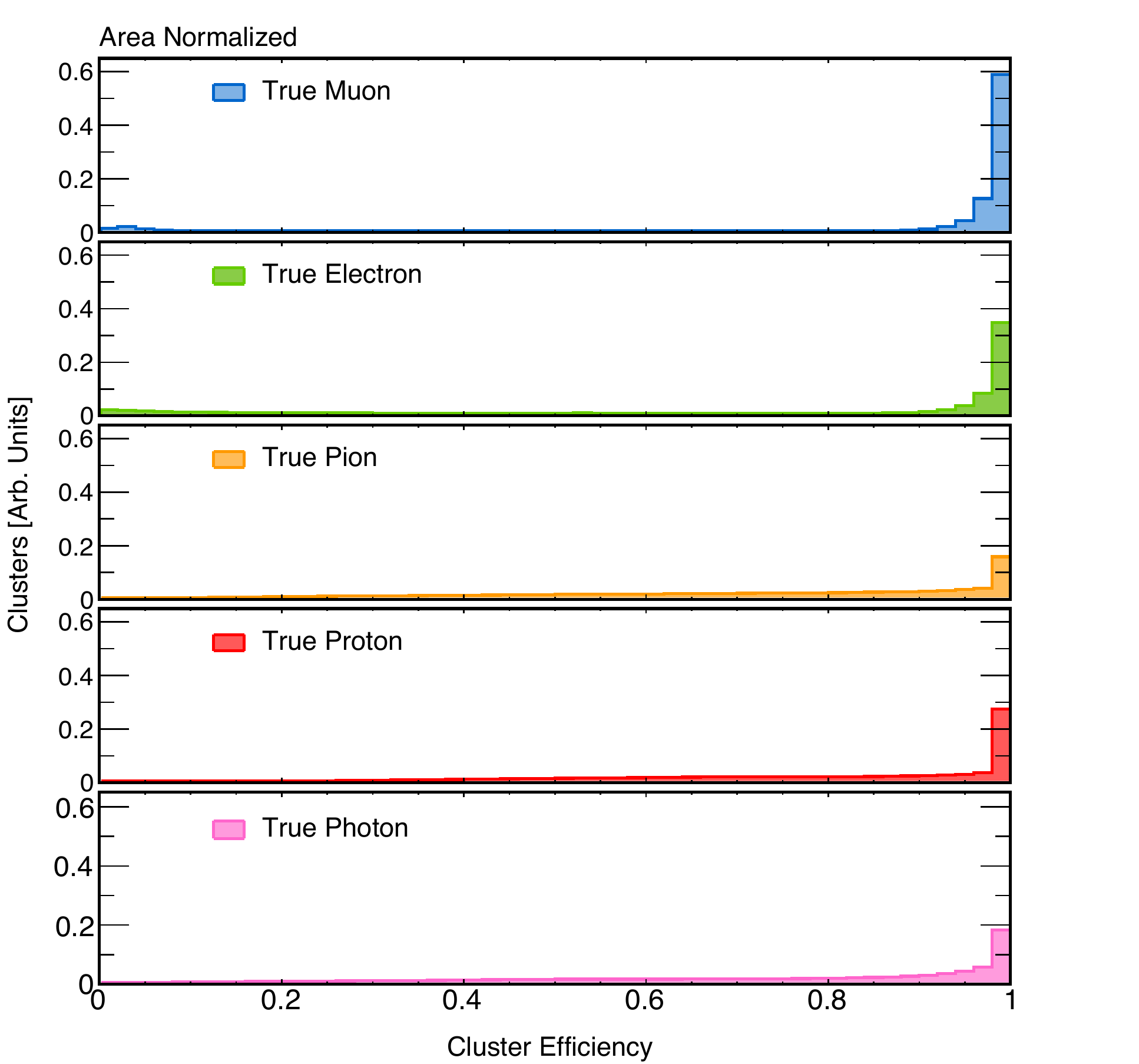}
     \includegraphics[width=0.49\linewidth,trim={270 75 15 0},clip=true]{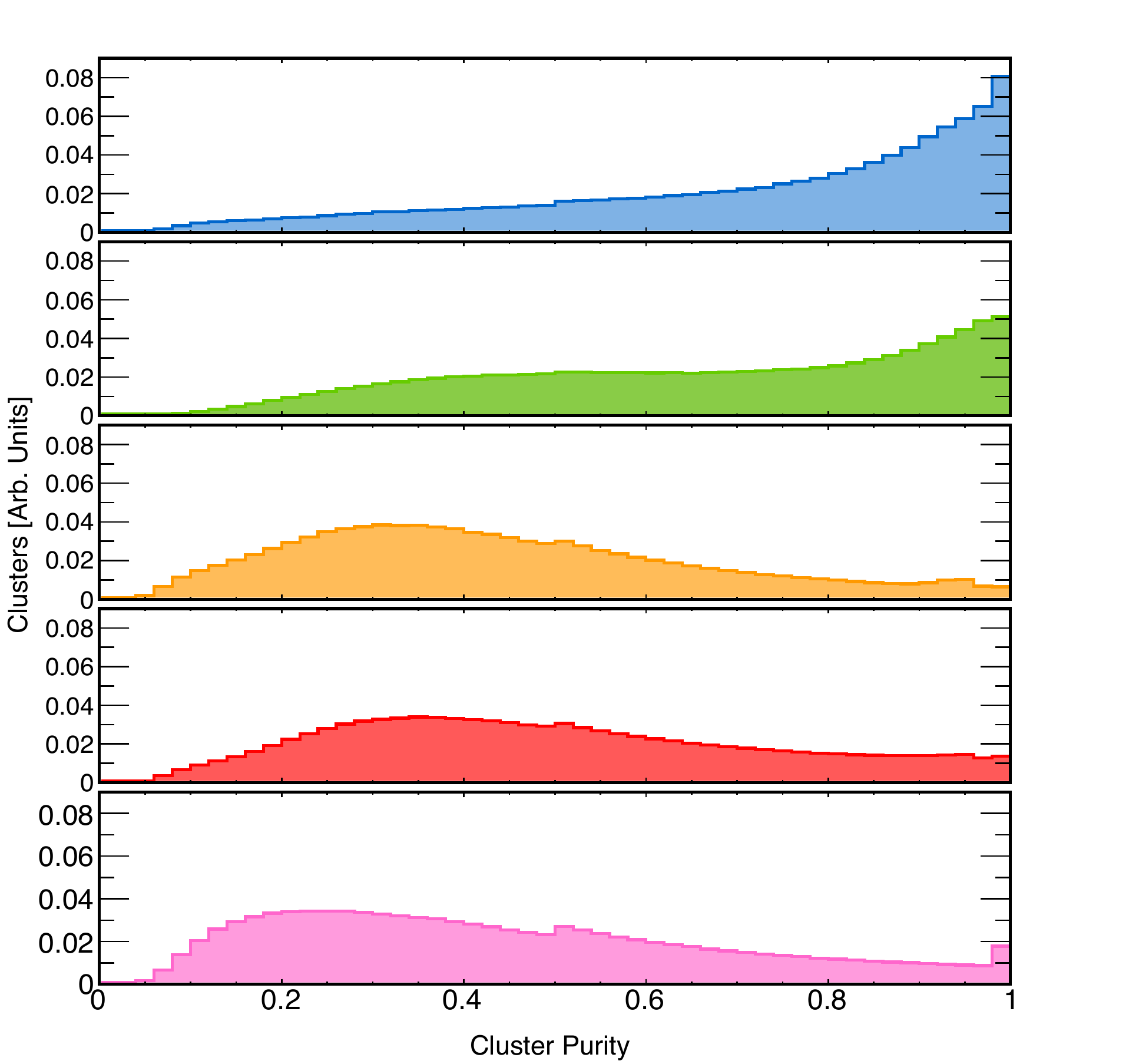}
     \caption{The efficiency and purity of reconstructed clusters in the NOvA detectors.
     Left: Cluster efficiency, defined as the fraction of energy depositions from the particle associated with a cluster which are contained by the cluster.
     Right: Cluster purity, defined as the fraction of the energy contained in a cluster which comes from the particle it is associated with.
     Clusters are associated with whichever particle deposited the majority of the energy contained in each cluster.
     A high efficiency indicates that clusters tend to mostly contain the energy deposits from the particle in question.
     The considerably lower purity results from the overlap of contributions from multiple particles in detector cells contained in one cluster.}
     \label{fig:prongPur}
\end{figure*}

A selection was applied to the events prior to training, requiring the ends of all clusters to be away from the edges of the detector, ensuring the use of fully contained events.
For each particle category, a threshold was chosen for the purity of the clusters, to balance a representative sample of realistic input data with clear identities of the clusters.
A purity of 0.5 or more was used for muons, electrons, and photons and a purity of 0.35 or more was used for pions and protons.
The efficiency and purity of all clusters for each type of particle can be seen in Figure~\ref{fig:prongPur}.

Only clusters shorter than five meters in length were used for training and evaluation to reduce total evaluation time on NOvA data.
Particle trajectories which are longer than 5~m come predominantly from muons (over 95\%), whose characteristic pattern of energy deposition can be easily identified with other techniques.

The final sample consisted of 2.95 million particles made up of 9\% muons, 33\% electrons, 9\% pions, 29\% protons, and 20\% photons.
There is no distinction in the sample between a particle and its antiparticle.
The sample was split, 80\% used for training and the remaining 20\% used to evaluate the algorithm's performance.

The network architecture 
is a simplified tower structure compared to GoogLeNet~\cite{googlenet} and to our previous implementation~\cite{eventCVN}.
The three consecutive inception modules per tower in the original architecture were reduced to one per tower~\cite{psihasthesis}, reducing the number of required computations with a roughly equivalent accuracy as the original classifier.
The kernel size was optimized for our running time requirements, and the width of the network was reduced by a factor of two, without significant losses in performance.
These improvements were also applied to the version of the Convolutional Visual Network (CVN) used by NOvA in its latest analysis~\cite{nova2018}.

\begin{figure}
     \centering
     \includegraphics[width=0.99\linewidth]{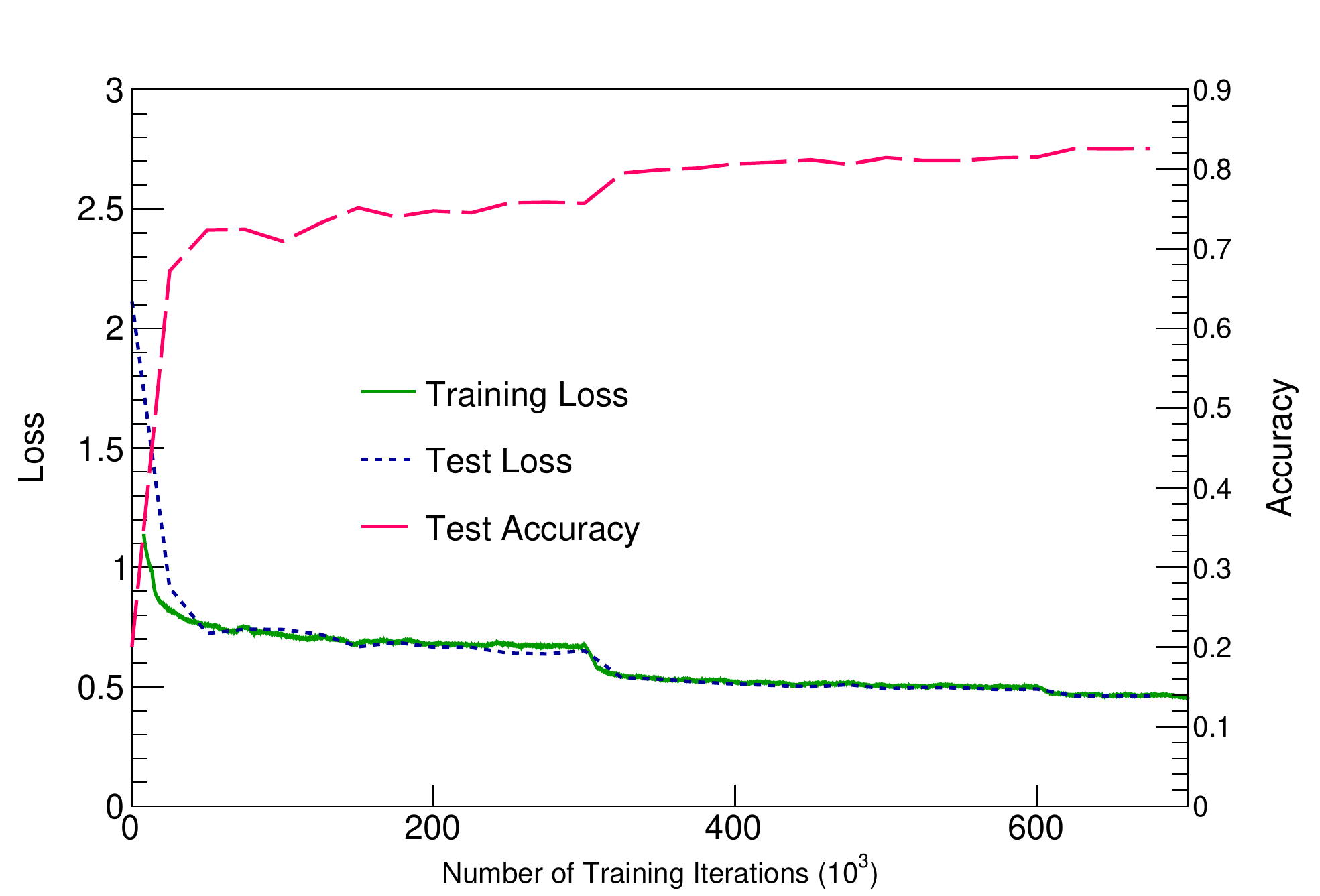}    
     \caption{The performance of the network as it is trained.
     The pink line depicts the accuracy of the network.
     The blue and green lines show the loss function of the test sample and training sample, respectively.}   
     \label{fig:loss}
\end{figure}

The model was optimized using stochastic gradient descent with categorical cross entropy as the loss function.
Figure~\ref{fig:loss} shows the loss and accuracy, a standard test of training stability.
The training and test losses agree indicating no signs of over-training.

The final layer of the network is a dense layer with five nodes, one for each training category, using a softmax activation function:
\begin{equation}
    \mathrm{S}\left(y_i\right) = \frac{\mathrm{e}^{y_i}}{\sum\limits_{j} \mathrm{e}^{y_j} }
\end{equation}
where $y_i$ are the outputs from the network before the activation function and S$\left(y_i\right)$ are the corresponding softmax scores, which yield a normalized set of outputs.

The network was trained using Caffe~\cite{caffe} 1.0 on four NVIDIA Tesla K20 GPUs on the BigRed II computing cluster at Indiana University.
The training spanned a total of 700,000 iterations with a batch size of 64.
After 300,000 iterations, the learning rate was reduced by a factor of 10.


Evaluation of NOvA data and simulations are produced in a ROOT~\cite{ROOT} format compatible with the \textit{art} framework~\cite{ART}.
Model evaluation was incorporated to the NOvA analysis infrastructure with the use of Caffe C++ API.
The source code and pretrained models for the two network architectures used in this work are available on \url{https://github.com/FernandaPsihas/ParticleTowerCNN}.

\section{Classification Performance}
\label{sec:performance}

\begin{figure*}
    \centering
    \includegraphics[width=0.99\linewidth]{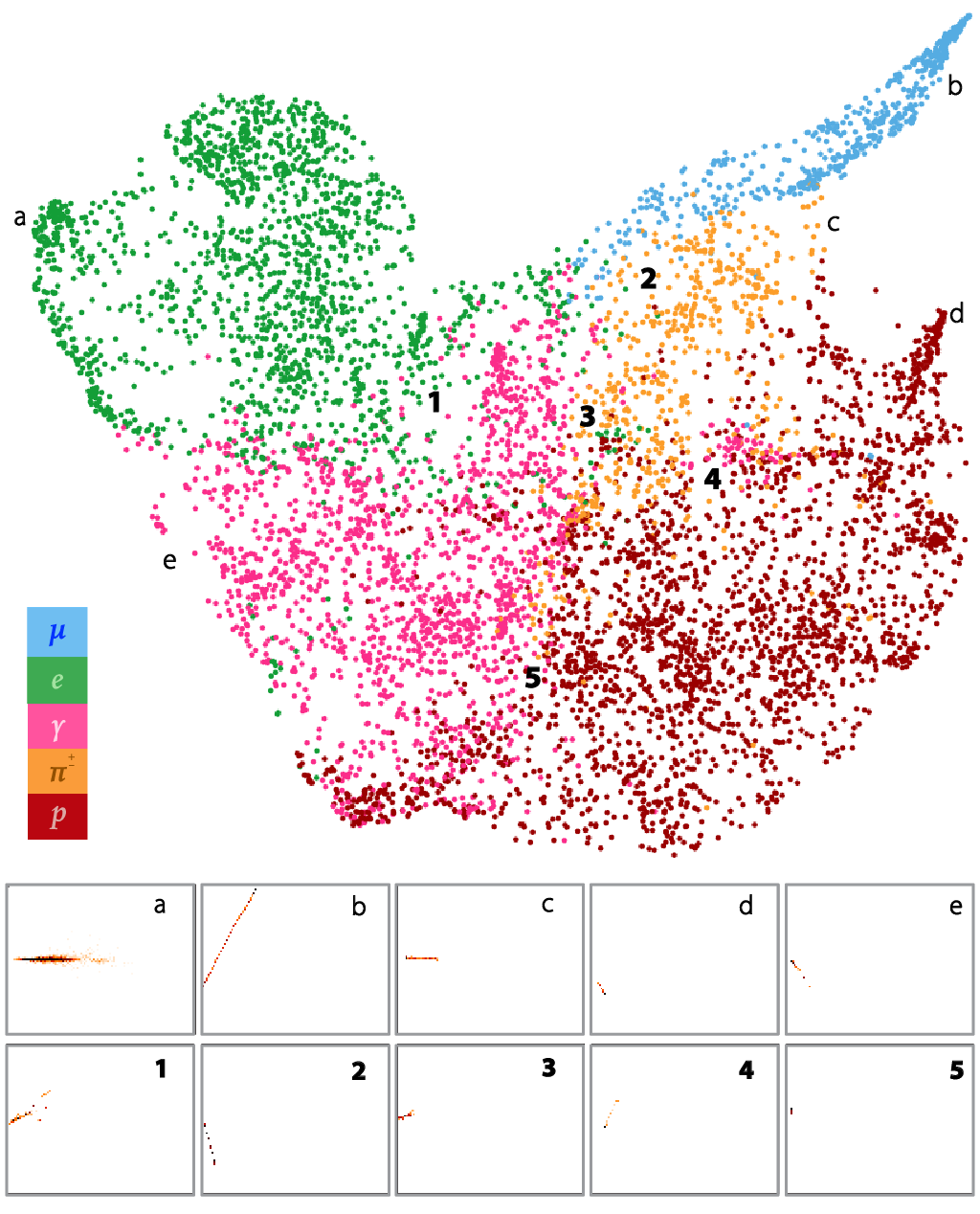}
    \caption{Classified events in two dimensional probability space created with the t-SNE~\cite{tsne} technique from the classifier features.
    A clear separation can be seen between muons (blue) and electrons (green).
    The expected overlap is visible between pions (orange) and protons (red).
    Regions with good separation and example topologies are labeled a-e.
    Regions with poor separation and example topologies are labeled 1-5}   
    \vspace{20pt}
    \label{fig:puppy}
\end{figure*}
The inputs to the network are two maps of the cluster without context and two maps with context each for the top and side view.
The full architecture is shown in Figure~\ref{fig:net}.
The weights are specific to each tower to encourage independent learning from each input.
The resulting features from the four towers are concatenated before a final inception, pooling, and fully connected layers.
To assess the effect of adding context to the input data, a particle-only network was trained using only the two views of the cluster for comparison.
The separation achieved for different particle types can be visualized from the two-dimensional mapping of events shown in Figure~\ref{fig:puppy}.
This mapping is produced using t-Distributed Stochastic Neighbor Embedding (t-SNE)~\cite{tsne}, a visualization algorithm which approximately maps similarity in high dimensional feature-space onto distance in two-dimensional space.
The t-SNE visualizations represent the overlap in extracted features as proximity in space, providing information regarding the discrimination power of the network.
Good separation is achieved for all particle types, but there are some expected regions of overlap, such as between electrons and photons.

\begin{figure*}
    \centering
    \begin{subfigure}[b]{0.48\textwidth}
        \centering
        \includegraphics[width=\textwidth]{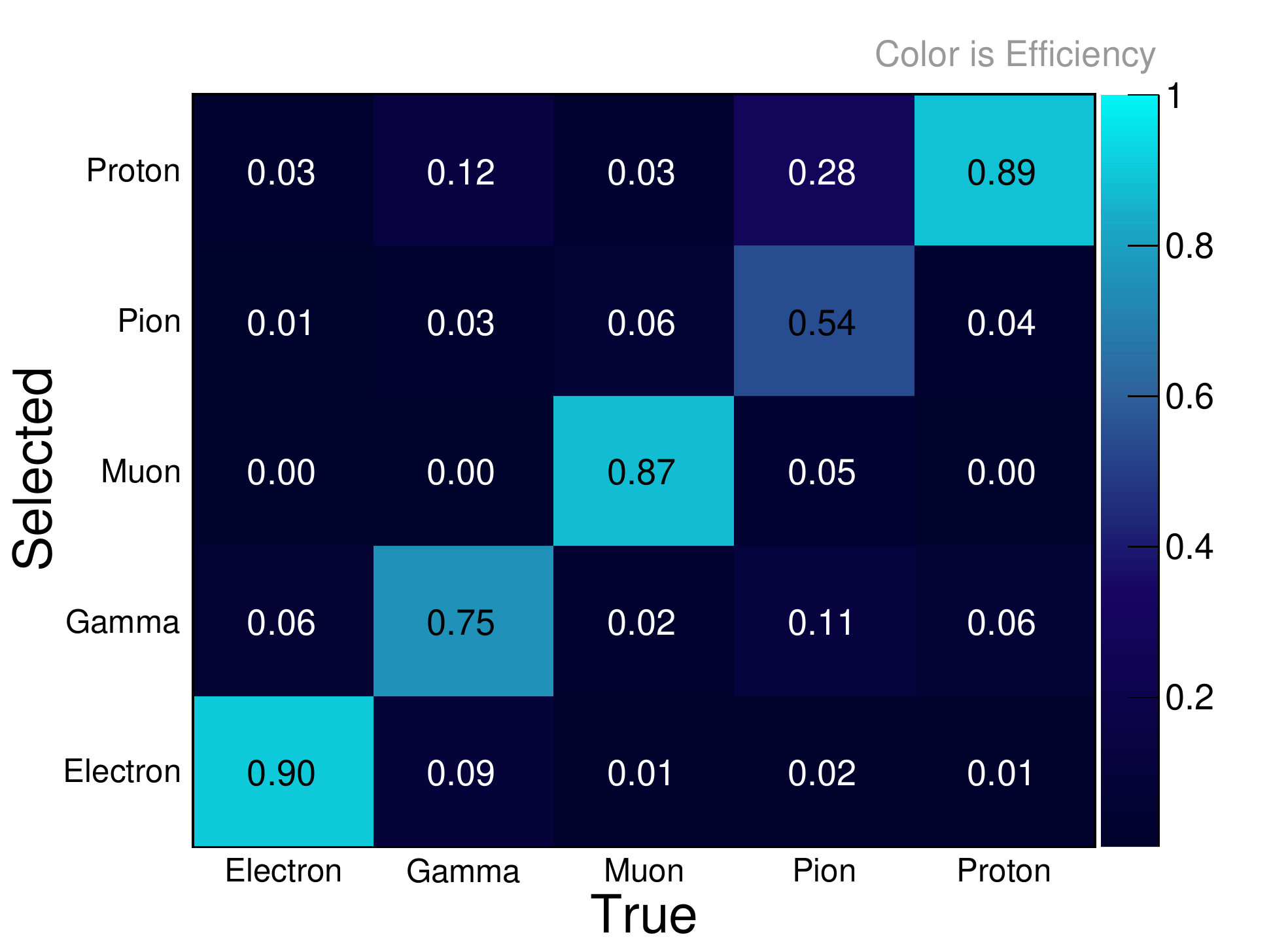}
    \end{subfigure}
    \hfill
    \begin{subfigure}[b]{0.48\textwidth}
        \centering
        \includegraphics[width=\textwidth]{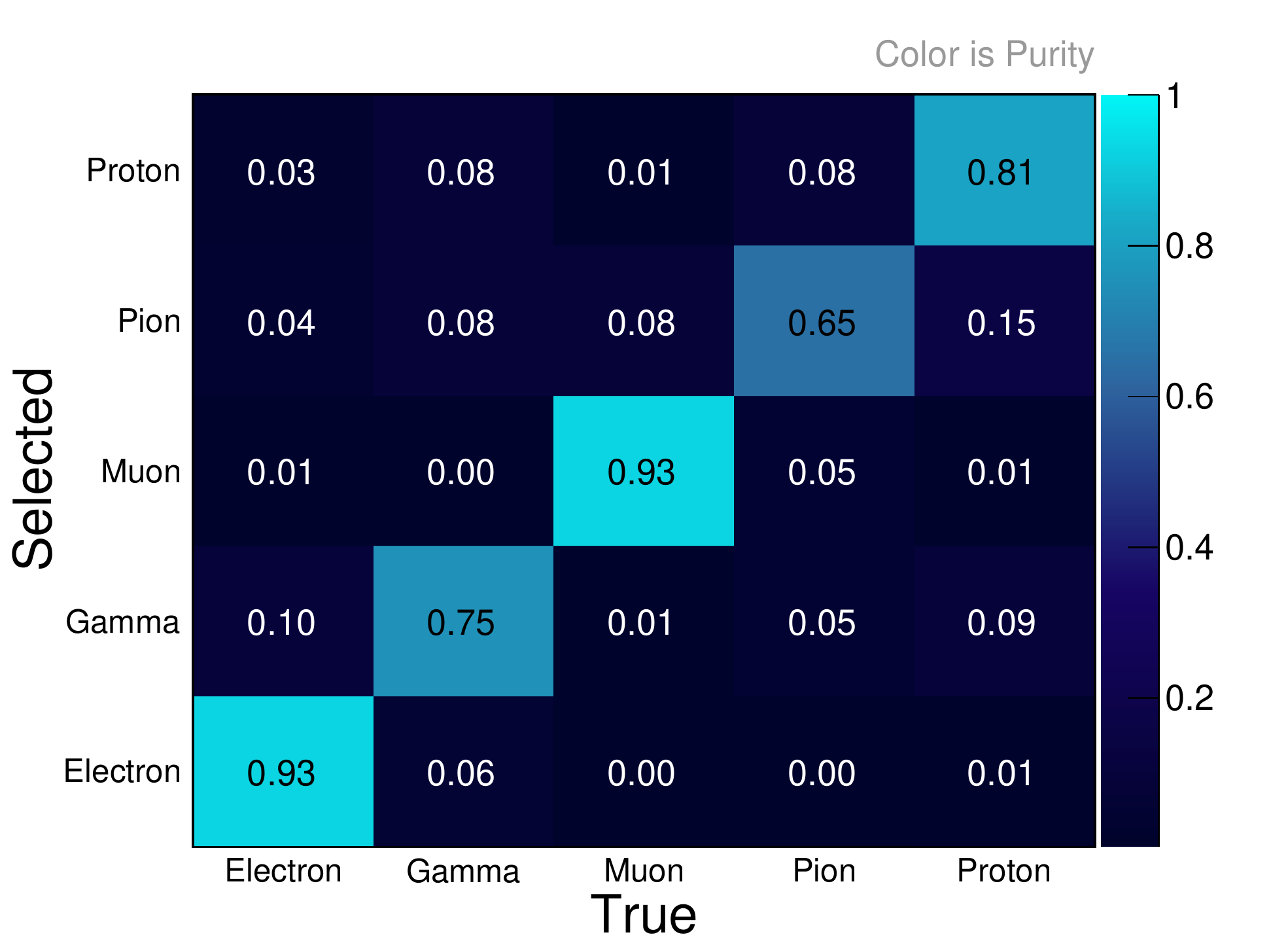}
    \end{subfigure}
    \begin{subfigure}[b]{0.48\textwidth}
        \centering
        \includegraphics[width=\textwidth]{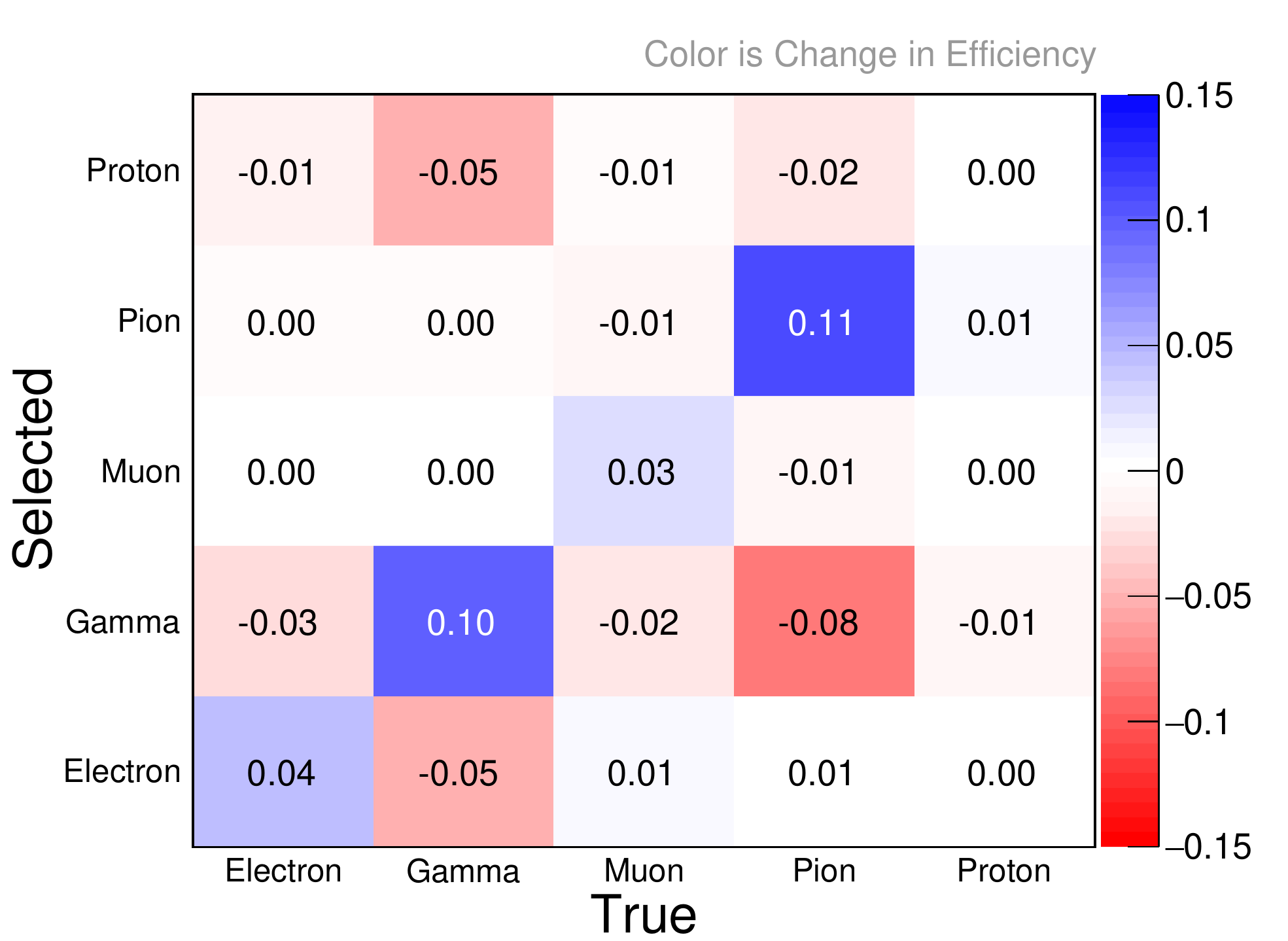}
    \end{subfigure}
    \hfill
    \begin{subfigure}[b]{0.48\textwidth}
        \centering
        \includegraphics[width=\textwidth]{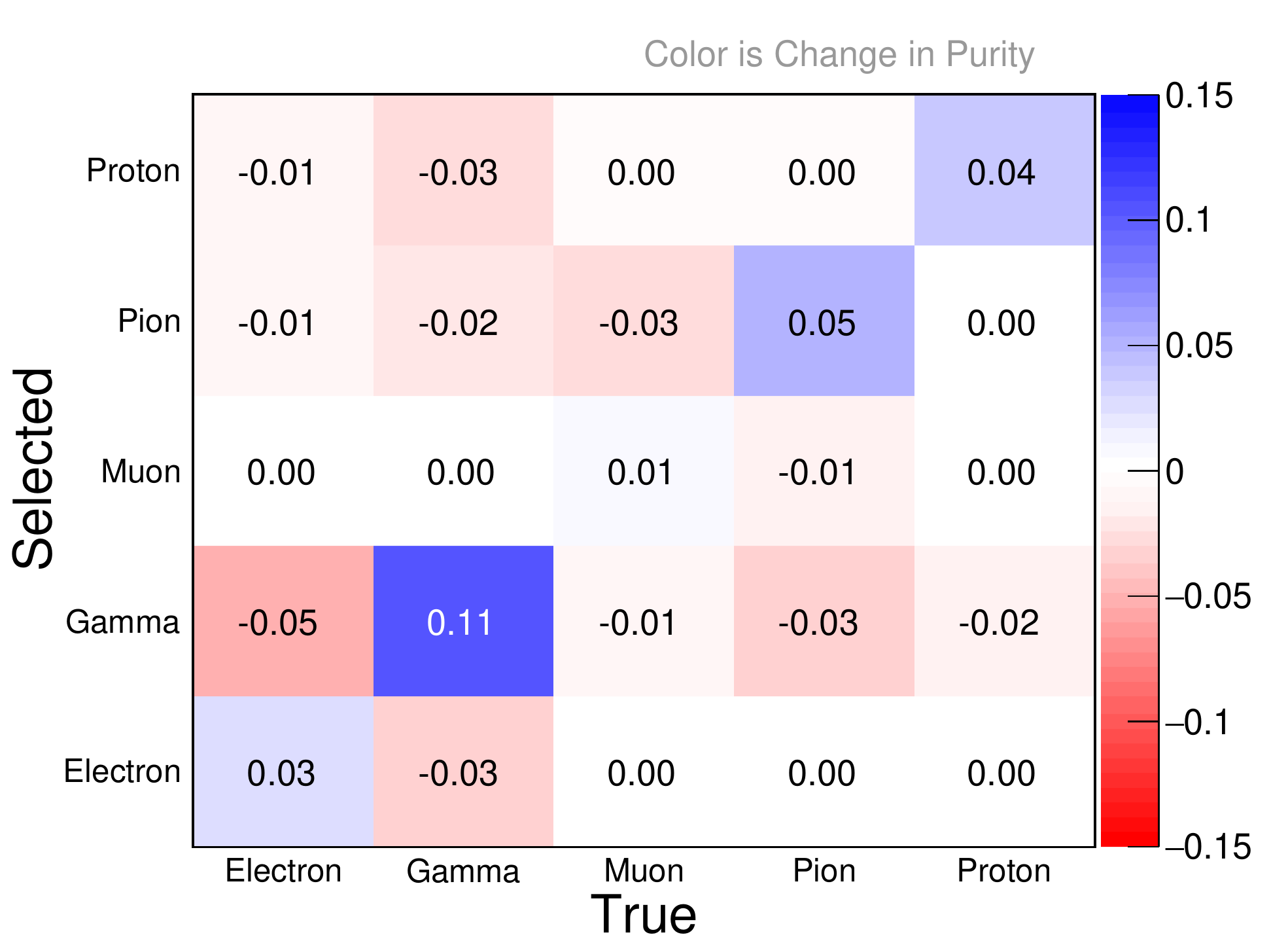}
    \end{subfigure}
    \caption{Classification matrices for the context-enriched network and its performance compared to the particle-only network.
    Top left: The classification matrix, column-normalized.
    The diagonal shows the efficiency for each category and the off-diagonal shows how events are misidentified.
    Top right: The classification matrix row normalized.
    The diagonal shows the purity for each category and the off-diagonal shows the backgrounds for each type of particle.
    Bottom: The change in efficiency (left) and purity (right) obtained from the subtraction of the particle only matrices from those shown above.
    The positive numbes on the diagonal elements show the absolute improvement in efficiency (left) and purity (right) obtained from adding context.}
    \label{fig:CM}
\end{figure*}

\begin{figure*}
    \centering
    \includegraphics[width=1\linewidth]{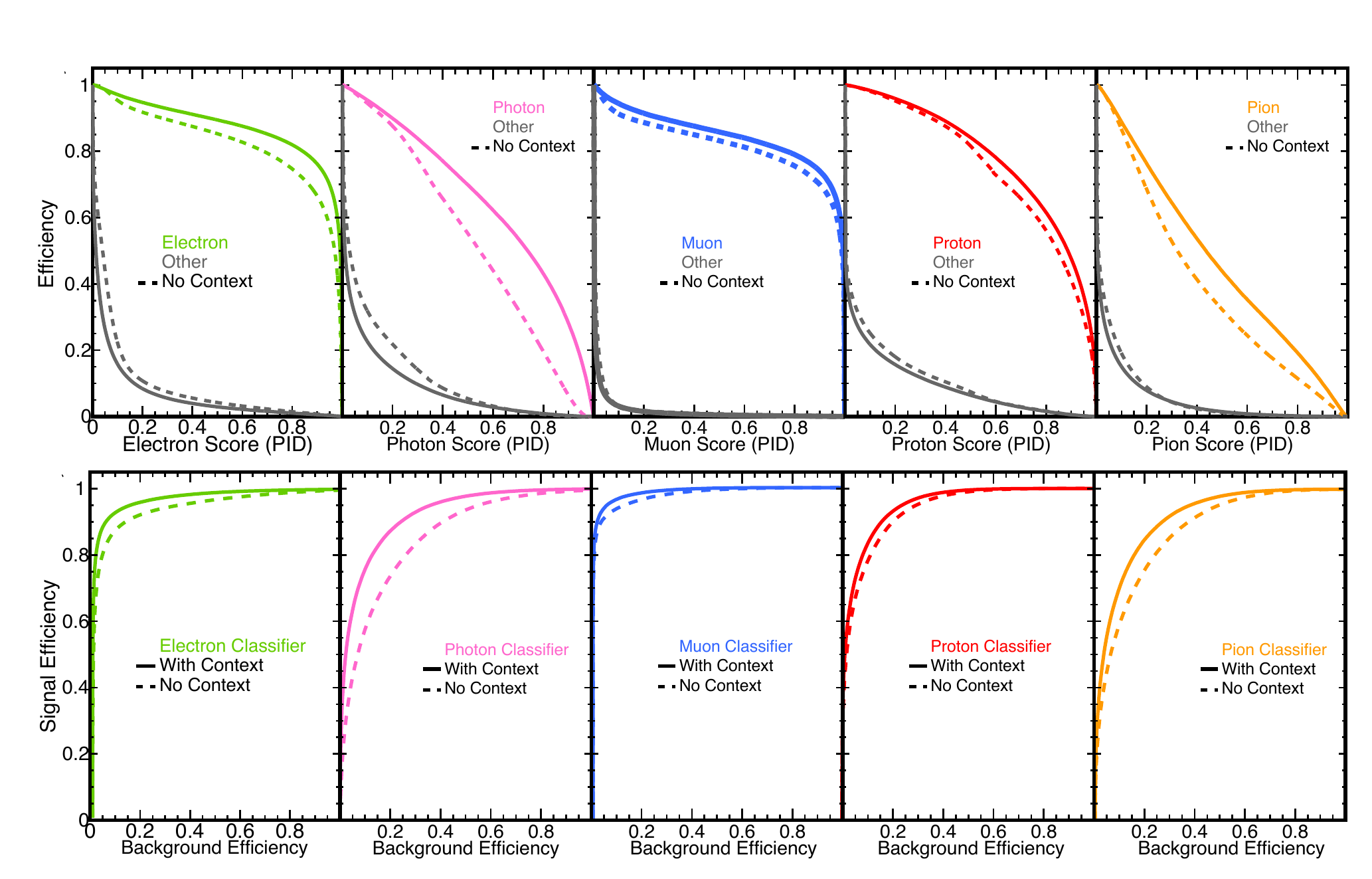}
    \caption{Selection efficiencies of individual softmax outputs for each training category taken as PID scores. Top: Signal and background selection efficiency vs PID score for each of the five particle classifiers with its respective background. Bottom: Signal vs background efficiency curves as the PID signal threshold is varied. The solid lines depict the efficiencies for the four-tower network utilizing the interaction context. The dashed lines depict the efficiencies for the two-tower network that only utilizes the particle cluster.
    }
    \label{fig:ROC-mu}
\end{figure*}

One class of metrics used for evaluating the classification takes the largest softmax value as the simple prediction from the network.
The top portion of Figure~\ref{fig:CM} shows the classification matrices for the networks simple prediction.
The network achieves 93\%, 75\%, 93\%, 65\%, and 81\% efficiency and 90\%, 75\%, 87\%, 54\%, and 89\% purity for electrons, photons, muons, pions, and protons without any additional selections.
The overall efficiency and purity of the classifier are 83.3\% and 83.5\% for all particles in the test sample.
The figure shows worse performance in the distinction between protons and pions as well as electrons and photons, which is expected given the differences noted in Section~\ref{sec:particles}.
We achieve separation of electrons and photons with only 6\% contamination of photons in the electron-selected events, and 10\% in the opposite case.
It is important to emphasize that only the largest score has been used for identification, Figure~\ref{fig:ROC-mu} indicates further potential to select particles with higher selection purity using all scores.

The network without context was trained on the same sample of events and employed the same network architecture, but using only two towers for both views of the reconstructed cluster.
The bottom portion of Figure~\ref{fig:CM} shows a comparison between the two networks.
In this case, the positive values along the diagonal quantify the improvement in efficiency or purity with inclusion of context in the network.
The improvement is more substantial for pions and photons, which gain up to 11\% in efficiency for the simple selection.
The photon-electron contamination improves by 5\%, for electrons identified as photons, and 3\%, for photons identified as electrons, and the misidentification of pions as photons is reduced by 8\% in the presence of context information.

Alternatively to the simple prediction based on the largest softmax value, the network outputs can also be used by considering each of the outputs as a Particle Identification (PID) score and by choosing a selection threshold optimized for a specific figure of merit such as efficiency, purity, etc.
Figure~\ref{fig:ROC-mu} 
shows the efficiencies of PID thresholds for each category. 
The performance on electrons and muons is superior to that for the other particles.
For photons and pions, good separation can be achieved with PID values above 0.7.

\begin{table*}
    \centering
    \begin{tabular}{l | l | c c c c c}
    \hline
    \hline
     \textbf{Comparison Metric} & \textbf{Network} & Electron & Photon & Muon & Pion & Proton \\
    \hline
    \multirow{2}{4.2cm}{Background Efficiency for 90\% Signal Efficiency} & Particle \& Context & 3.2\% & 14.5\% & 1.1\% & 16.1\% & 9.4\% \\
     & Particle Only & 8.0\% & 24.6\% & 2.2\% & 22.4\% & 12.1\% \\
    \hline
    \multirow{2}{*}{ROC Integral} & Particle \& Context & 0.98 & 0.95 & 0.99 & 0.94 & 0.97 \\
     & Particle Only & 0.97 & 0.91 & 0.99 & 0.92 & 0.96 \\
    \hline
    \multirow{2}{3.3cm}{Largest Score Selection Efficiency} & Particle \& Context & 90\% & 75\% & 87\% & 54\% & 89\% \\
     & Particle Only & 86\% & 74\% & 84\% & 43\% & 89\% \\
    \hline
    \multirow{2}{3.3cm}{Largest Score Selection Purity} & Particle \& Context & 93\% & 75\% & 93\% & 65\% & 81\% \\
     & Particle Only & 90\% & 64\% & 92\% & 60\% & 77\% \\
    \hline
    \hline
    \end{tabular}
    \caption{Summary comparisons of the particle with context and the particle only networks.
    The largest score selection is the simple softmax selection shown in Figure~\ref{fig:CM}.
    The ROC integrals and background efficiencies were obtained from Figure~\ref{fig:ROC-mu}.}
    \label{tab:roc_tables}
\end{table*}

The bottom portion of Figure~\ref{fig:ROC-mu} also shows the Receiver Operating Characteristic (ROC) curves for the classification of each particle type.
ROC curves show the discriminating power of the classifier when taking the softmax output as a PID score and by varying the PID signal selection threshold between 0 to 1.
Separation between the curves corresponding to each training scheme indicates a  difference in performance in terms of signal vs background efficiency.
The classifier which uses context information outperforms the particle-only classifier  in every category, though the difference is most significant for photon, pion, and electron identification. 

Table~\ref{tab:roc_tables} contains a summary of the metrics used to compare the network trained using context and the particle-only network. These results clearly show the improvement that adding context information provides to the task of classification.

More detailed studies are ongoing to assess opportunities for improvement from generation and selection of events for the training sample, bias reduction, and treatment of systematic uncertainties.
Thus far, we have found no evidence of bias coming from simulated particle kinematics.
Studies of detector response have also been implemented using this technique for the selection of photon pairs consistent with coming from neutral pion decays.
The event selection using this technique achieves a purity of 92\%, an improvement of 60\% for the same efficiency compared to previous methods~\cite{psihasthesis}.

This context-enriched CNN technique for particle classification has already been incorporated in NOvA's analyses, specifically in reconstructing the energy of electron neutrino interactions~\cite{psihasthesis}, an essential part of NOvA's primary neutrino oscillation physics measurement. The energy in electron neutrino candidate events is reconstructed from electromagnetic activity by final-state electrons and photons, and from hadronic activity by final-state protons and pions.
Each cluster is identified as 
electromagnetic or hadronic using this classifier.
The energy resolution using this method is 11\%~\cite{nova2018}, an improvement of 20\% compared to previous methods.
This context-enriched CNN technique is being explored further for selection of interaction final states~\cite{neutrinoNuNC} as well as energy reconstruction for muon neutrino candidate events~\cite{neutrinoEnergy}.

\section{Conclusion}

This work describes the first implementation of a deep learning algorithm which introduces interaction context to the problem of particle classification, where the context is introduced through the use of a four-tower, siamese, convolutional neural network architecture.
We demonstrate the first application of the particle classifier based on this architecture to neutrino detector data.
The overall efficiency and purity of the classifier for all particles in the test sample are 83.3\% and 83.5\%, respectively.
The addition of context improves the performance of the classifier in all metrics, particularly in the categories most expected to benefit from contextual information.
Improvements were found in both efficiency and purity for all particle types, in particular an 11\% increase in the efficiency for simple selection (largest softmax value) of pions.

Detailed studies are ongoing to assess opportunities for improvement in sample composition, bias reduction, and treatment of systematic uncertainties.
In particular, studies involving tagged data from the upcoming NOvA test-beam program~\cite{testbeam}, and comparative studies with single particle simulations are being developed.
Future studies in applications of CNNs to physics analyses are continuing, with focus on better understanding the impact of the simulated training samples in their performance and for analysis of particle data.

\section*{Acknowledgements}
We thank the NOvA collaboration for the use of its software tools and simulation and for the review of this manuscript. This work was supported by the U.S. Department of Energy and the U.S. National Science Foundation.
Fernanda Psihas's contributions were funded in part by CONACyT and the University of Texas system through the ConTex program.
NOvA receives additional support from the Department of Science and Technology, India;
the European Research Council; the MSMT CR, Czech Republic; the RAS, RMES, and RFBR,
Russia; CNPq and FAPEG, Brazil; and the State and University of Minnesota.
We are grateful for the contributions of the staff at the Ash River Laboratory, Argonne National Laboratory, and Fermilab.
Fermilab is operated by Fermi Research Alliance, LLC under Contract No.De-AC02-07CH11359 with the U.S. DOE.

We are grateful to the Fermilab Scientific Computing Division for support and maintenance of the GPU cluster which was used in early stages of this work.

This research was supported in part by Lilly Endowment, Inc., through its support for the Indiana University Pervasive Technology Institute, and in part by the Indiana METACyt Initiative. The Indiana METACyt Initiative at IU was also supported in part by Lilly Endowment, Inc.

\bibliography{main}

\begin{thebibliography}{34}%
\makeatletter
\providecommand \@ifxundefined [1]{%
 \@ifx{#1\undefined}
}%
\providecommand \@ifnum [1]{%
 \ifnum #1\expandafter \@firstoftwo
 \else \expandafter \@secondoftwo
 \fi
}%
\providecommand \@ifx [1]{%
 \ifx #1\expandafter \@firstoftwo
 \else \expandafter \@secondoftwo
 \fi
}%
\providecommand \natexlab [1]{#1}%
\providecommand \enquote  [1]{``#1''}%
\providecommand \bibnamefont  [1]{#1}%
\providecommand \bibfnamefont [1]{#1}%
\providecommand \citenamefont [1]{#1}%
\providecommand \href@noop [0]{\@secondoftwo}%
\providecommand \href [0]{\begingroup \@sanitize@url \@href}%
\providecommand \@href[1]{\@@startlink{#1}\@@href}%
\providecommand \@@href[1]{\endgroup#1\@@endlink}%
\providecommand \@sanitize@url [0]{\catcode `\\12\catcode `\$12\catcode
  `\&12\catcode `\#12\catcode `\^12\catcode `\_12\catcode `\%12\relax}%
\providecommand \@@startlink[1]{}%
\providecommand \@@endlink[0]{}%
\providecommand \url  [0]{\begingroup\@sanitize@url \@url }%
\providecommand \@url [1]{\endgroup\@href {#1}{\urlprefix }}%
\providecommand \urlprefix  [0]{URL }%
\providecommand \Eprint [0]{\href }%
\providecommand \doibase [0]{https://doi.org/}%
\providecommand \selectlanguage [0]{\@gobble}%
\providecommand \bibinfo  [0]{\@secondoftwo}%
\providecommand \bibfield  [0]{\@secondoftwo}%
\providecommand \translation [1]{[#1]}%
\providecommand \BibitemOpen [0]{}%
\providecommand \bibitemStop [0]{}%
\providecommand \bibitemNoStop [0]{.\EOS\space}%
\providecommand \EOS [0]{\spacefactor3000\relax}%
\providecommand \BibitemShut  [1]{\csname bibitem#1\endcsname}%
\let\auto@bib@innerbib\@empty
\bibitem [{\citenamefont {Glaser}\ and\ \citenamefont {Rahm}(1955)}]{bubbles}%
  \BibitemOpen
  \bibfield  {author} {\bibinfo {author} {\bibfnamefont {D.~A.}\ \bibnamefont
  {Glaser}}and\ \bibinfo {author} {\bibfnamefont {D.~C.}\ \bibnamefont
  {Rahm}},\ }\bibfield  {title} {\bibinfo {title} {Characteristics of bubble
  chambers},\ }\href {https://doi.org/10.1103/PhysRev.97.474} {\bibfield
  {journal} {\bibinfo  {journal} {Phys. Rev.}\ }\textbf {\bibinfo {volume}
  {97}},\ \bibinfo {pages} {474} (\bibinfo {year} {1955})}\BibitemShut
  {NoStop}%
\bibitem [{\citenamefont {Radovic}\ \emph {et~al.}(2018)\citenamefont
  {Radovic}, \citenamefont {Williams}, \citenamefont {Rousseau}, \citenamefont
  {Kagan}, \citenamefont {Bonacorsi}, \citenamefont {Himmel}, \citenamefont
  {Aurisano}, \citenamefont {Terao},\ and\ \citenamefont {Wongjirad}}]{Nature}%
  \BibitemOpen
  \bibfield  {author} {\bibinfo {author} {\bibfnamefont {A.}~\bibnamefont
  {Radovic}}, \bibinfo {author} {\bibfnamefont {M.}~\bibnamefont {Williams}},
  \bibinfo {author} {\bibfnamefont {D.}~\bibnamefont {Rousseau}}, \bibinfo
  {author} {\bibfnamefont {M.}~\bibnamefont {Kagan}}, \bibinfo {author}
  {\bibfnamefont {D.}~\bibnamefont {Bonacorsi}}, \bibinfo {author}
  {\bibfnamefont {A.}~\bibnamefont {Himmel}}, \bibinfo {author} {\bibfnamefont
  {A.}~\bibnamefont {Aurisano}}, \bibinfo {author} {\bibfnamefont
  {K.}~\bibnamefont {Terao}}, and\ \bibinfo {author} {\bibfnamefont
  {T.}~\bibnamefont {Wongjirad}},\ }\bibfield  {title} {\bibinfo {title}
  {{Machine learning at the energy and intensity frontiers of particle
  physics}},\ }\href {https://doi.org/10.1038/s41586-018-0361-2} {\bibfield
  {journal} {\bibinfo  {journal} {Nature}\ }\textbf {\bibinfo {volume} {560}},\
  \bibinfo {pages} {41} (\bibinfo {year} {2018})}\BibitemShut {NoStop}%
\bibitem [{\citenamefont {de~Oliveira}\ \emph {et~al.}(2016)\citenamefont
  {de~Oliveira}, \citenamefont {Kagan}, \citenamefont {Mackey}, \citenamefont
  {Nachman},\ and\ \citenamefont {Schwartzman}}]{kagan}%
  \BibitemOpen
  \bibfield  {author} {\bibinfo {author} {\bibfnamefont {L.}~\bibnamefont
  {de~Oliveira}}, \bibinfo {author} {\bibfnamefont {M.}~\bibnamefont {Kagan}},
  \bibinfo {author} {\bibfnamefont {L.}~\bibnamefont {Mackey}}, \bibinfo
  {author} {\bibfnamefont {B.}~\bibnamefont {Nachman}}, and\ \bibinfo {author}
  {\bibfnamefont {A.}~\bibnamefont {Schwartzman}},\ }\bibfield  {title}
  {\bibinfo {title} {{Jet-images - deep learning edition}},\ }\href
  {https://doi.org/10.1007/JHEP07(2016)069} {\bibfield  {journal} {\bibinfo
  {journal} {JHEP}\ }\textbf {\bibinfo {volume} {07}},\ \bibinfo {pages}
  {069}},\ \Eprint {https://arxiv.org/abs/1511.05190} {arXiv:1511.05190
  [hep-ph]} \BibitemShut {NoStop}%
\bibitem [{\citenamefont {Aurisano}\ \emph {et~al.}(2016)\citenamefont
  {Aurisano}, \citenamefont {Radovic}, \citenamefont {Rocco}, \citenamefont
  {Himmel}, \citenamefont {Messier}, \citenamefont {Niner}, \citenamefont
  {Pawloski}, \citenamefont {Psihas}, \citenamefont {Sousa},\ and\
  \citenamefont {Vahle}}]{eventCVN}%
  \BibitemOpen
  \bibfield  {author} {\bibinfo {author} {\bibfnamefont {A.}~\bibnamefont
  {Aurisano}}, \bibinfo {author} {\bibfnamefont {A.}~\bibnamefont {Radovic}},
  \bibinfo {author} {\bibfnamefont {D.}~\bibnamefont {Rocco}}, \bibinfo
  {author} {\bibfnamefont {A.}~\bibnamefont {Himmel}}, \bibinfo {author}
  {\bibfnamefont {M.~D.}\ \bibnamefont {Messier}}, \bibinfo {author}
  {\bibfnamefont {E.}~\bibnamefont {Niner}}, \bibinfo {author} {\bibfnamefont
  {G.}~\bibnamefont {Pawloski}}, \bibinfo {author} {\bibfnamefont
  {F.}~\bibnamefont {Psihas}}, \bibinfo {author} {\bibfnamefont
  {A.}~\bibnamefont {Sousa}}, and\ \bibinfo {author} {\bibfnamefont
  {P.}~\bibnamefont {Vahle}},\ }\bibfield  {title} {\bibinfo {title} {{A
  Convolutional Neural Network Neutrino Event Classifier}},\ }\href
  {https://doi.org/10.1088/1748-0221/11/09/P09001} {\bibfield  {journal}
  {\bibinfo  {journal} {JINST}\ }\textbf {\bibinfo {volume} {11}}\bibfield
  {number} {\bibinfo  {number} { (09)},\ \bibinfo {pages} {P09001}},\ }\Eprint
  {https://arxiv.org/abs/1604.01444} {arXiv:1604.01444 [hep-ex]} \BibitemShut
  {NoStop}%
\bibitem [{\citenamefont {Paganini}\ \emph {et~al.}(2018)\citenamefont
  {Paganini}, \citenamefont {de~Oliveira},\ and\ \citenamefont
  {Nachman}}]{GANs}%
  \BibitemOpen
  \bibfield  {author} {\bibinfo {author} {\bibfnamefont {M.}~\bibnamefont
  {Paganini}}, \bibinfo {author} {\bibfnamefont {L.}~\bibnamefont
  {de~Oliveira}}, and\ \bibinfo {author} {\bibfnamefont {B.}~\bibnamefont
  {Nachman}},\ }\bibfield  {title} {\bibinfo {title} {{CaloGAN : Simulating 3D
  high energy particle showers in multilayer electromagnetic calorimeters with
  generative adversarial networks}},\ }\href
  {https://doi.org/10.1103/PhysRevD.97.014021} {\bibfield  {journal} {\bibinfo
  {journal} {Phys. Rev.}\ }\textbf {\bibinfo {volume} {D97}},\ \bibinfo {pages}
  {014021} (\bibinfo {year} {2018})},\ \Eprint
  {https://arxiv.org/abs/1712.10321} {arXiv:1712.10321 [hep-ex]} \BibitemShut
  {NoStop}%
\bibitem [{\citenamefont {Baldi}\ \emph {et~al.}(2018)\citenamefont {Baldi},
  \citenamefont {Bian}, \citenamefont {Hertel},\ and\ \citenamefont
  {Li}}]{JMEnergy}%
  \BibitemOpen
  \bibfield  {author} {\bibinfo {author} {\bibfnamefont {P.}~\bibnamefont
  {Baldi}}, \bibinfo {author} {\bibfnamefont {J.}~\bibnamefont {Bian}},
  \bibinfo {author} {\bibfnamefont {L.}~\bibnamefont {Hertel}}, and\ \bibinfo
  {author} {\bibfnamefont {L.}~\bibnamefont {Li}},\ }\bibfield  {title}
  {\bibinfo {title} {{Improved Energy Reconstruction in NOvA with Regression
  Convolutional Neural Networks}},\ }\href@noop {} {\  (\bibinfo {year}
  {2018})},\ \Eprint {https://arxiv.org/abs/1811.04557} {arXiv:1811.04557
  [physics.ins-det]} \BibitemShut {NoStop}%
\bibitem [{\citenamefont {Acciarri}\ \emph {et~al.}(2017)\citenamefont
  {Acciarri} \emph {et~al.}}]{microboone}%
  \BibitemOpen
  \bibfield  {author} {\bibinfo {author} {\bibfnamefont {R.}~\bibnamefont
  {Acciarri}} \emph {et~al.} (\bibinfo {collaboration} {MicroBooNE}),\
  }\bibfield  {title} {\bibinfo {title} {{Convolutional Neural Networks Applied
  to Neutrino Events in a Liquid Argon Time Projection Chamber}},\ }\href
  {https://doi.org/10.1088/1748-0221/12/03/P03011} {\bibfield  {journal}
  {\bibinfo  {journal} {JINST}\ }\textbf {\bibinfo {volume} {12}}\bibfield
  {number} {\bibinfo  {number} { (03)},\ \bibinfo {pages} {P03011}},\ }\Eprint
  {https://arxiv.org/abs/1611.05531} {arXiv:1611.05531 [physics.ins-det]}
  \BibitemShut {NoStop}%
\bibitem [{\citenamefont {Renner}\ \emph {et~al.}(2017)\citenamefont {Renner}
  \emph {et~al.}}]{next}%
  \BibitemOpen
  \bibfield  {author} {\bibinfo {author} {\bibfnamefont {J.}~\bibnamefont
  {Renner}} \emph {et~al.} (\bibinfo {collaboration} {NEXT}),\ }\bibfield
  {title} {\bibinfo {title} {{Background rejection in NEXT using deep neural
  networks}},\ }\href {https://doi.org/10.1088/1748-0221/12/01/T01004}
  {\bibfield  {journal} {\bibinfo  {journal} {JINST}\ }\textbf {\bibinfo
  {volume} {12}}\bibfield  {number} {\bibinfo  {number} { (01)},\ \bibinfo
  {pages} {T01004}},\ }\Eprint {https://arxiv.org/abs/1609.06202}
  {arXiv:1609.06202 [physics.ins-det]} \BibitemShut {NoStop}%
\bibitem [{\citenamefont {Perdue}\ \emph {et~al.}(2018)\citenamefont {Perdue}
  \emph {et~al.}}]{MinervaVTX}%
  \BibitemOpen
  \bibfield  {author} {\bibinfo {author} {\bibfnamefont {G.~N.}\ \bibnamefont
  {Perdue}} \emph {et~al.} (\bibinfo {collaboration} {MINERvA}),\ }\bibfield
  {title} {\bibinfo {title} {{Reducing model bias in a deep learning classifier
  using domain adversarial neural networks in the MINERvA experiment}},\ }\href
  {https://doi.org/10.1088/1748-0221/13/11/P11020} {\bibfield  {journal}
  {\bibinfo  {journal} {JINST}\ }\textbf {\bibinfo {volume} {13}}\bibfield
  {number} {\bibinfo  {number} { (11)},\ \bibinfo {pages} {P11020}},\ }\Eprint
  {https://arxiv.org/abs/1808.08332} {arXiv:1808.08332 [physics.data-an]}
  \BibitemShut {NoStop}%
\bibitem [{\citenamefont {Adams}\ \emph {et~al.}(2018)\citenamefont {Adams}
  \emph {et~al.}}]{microbooneSS}%
  \BibitemOpen
  \bibfield  {author} {\bibinfo {author} {\bibfnamefont {C.}~\bibnamefont
  {Adams}} \emph {et~al.} (\bibinfo {collaboration} {MicroBooNE}),\ }\bibfield
  {title} {\bibinfo {title} {{A Deep Neural Network for Pixel-Level
  Electromagnetic Particle Identification in the MicroBooNE Liquid Argon Time
  Projection Chamber}},\ }\href@noop {} {\  (\bibinfo {year} {2018})},\ \Eprint
  {https://arxiv.org/abs/1808.07269} {arXiv:1808.07269 [physics.ins-det]}
  \BibitemShut {NoStop}%
\bibitem [{\citenamefont {Acero}\ \emph {et~al.}(2018)\citenamefont {Acero}
  \emph {et~al.}}]{nova2018}%
  \BibitemOpen
  \bibfield  {author} {\bibinfo {author} {\bibfnamefont {M.~A.}\ \bibnamefont
  {Acero}} \emph {et~al.} (\bibinfo {collaboration} {NOvA}),\ }\bibfield
  {title} {\bibinfo {title} {{New constraints on oscillation parameters from
  $\nu_e$ appearance and $\nu_\mu$ disappearance in the NOvA experiment}},\
  }\href {https://doi.org/10.1103/PhysRevD.98.032012} {\bibfield  {journal}
  {\bibinfo  {journal} {Phys. Rev.}\ }\textbf {\bibinfo {volume} {D98}},\
  \bibinfo {pages} {032012} (\bibinfo {year} {2018})},\ \Eprint
  {https://arxiv.org/abs/1806.00096} {arXiv:1806.00096 [hep-ex]} \BibitemShut
  {NoStop}%
\bibitem [{\citenamefont {Katori}\ and\ \citenamefont
  {Martini}(2018)}]{nuintreview}%
  \BibitemOpen
  \bibfield  {author} {\bibinfo {author} {\bibfnamefont {T.}~\bibnamefont
  {Katori}}and\ \bibinfo {author} {\bibfnamefont {M.}~\bibnamefont {Martini}},\
  }\bibfield  {title} {\bibinfo {title} {{Neutrino-nucleus cross sections for
  oscillation experiments}},\ }\href {https://doi.org/10.1088/1361-6471/aa8bf7}
  {\bibfield  {journal} {\bibinfo  {journal} {J. Phys.}\ }\textbf {\bibinfo
  {volume} {G45}},\ \bibinfo {pages} {013001} (\bibinfo {year} {2018})},\
  \Eprint {https://arxiv.org/abs/1611.07770} {arXiv:1611.07770 [hep-ph]}
  \BibitemShut {NoStop}%
\bibitem [{\citenamefont {Bromley}\ \emph {et~al.}(1993)\citenamefont
  {Bromley}, \citenamefont {Guyon}, \citenamefont {LeCun}, \citenamefont
  {S\"{a}ckinger},\ and\ \citenamefont {Shah}}]{siamese}%
  \BibitemOpen
  \bibfield  {author} {\bibinfo {author} {\bibfnamefont {J.}~\bibnamefont
  {Bromley}}, \bibinfo {author} {\bibfnamefont {I.}~\bibnamefont {Guyon}},
  \bibinfo {author} {\bibfnamefont {Y.}~\bibnamefont {LeCun}}, \bibinfo
  {author} {\bibfnamefont {E.}~\bibnamefont {S\"{a}ckinger}}, and\ \bibinfo
  {author} {\bibfnamefont {R.}~\bibnamefont {Shah}},\ }\bibfield  {title}
  {\bibinfo {title} {Signature verification using a "siamese" time delay neural
  network},\ }in\ \href {http://dl.acm.org/citation.cfm?id=2987189.2987282}
  {\emph {\bibinfo {booktitle} {Proceedings of the 6th International Conference
  on Neural Information Processing Systems}}},\ \bibinfo {series and number}
  {NIPS'93}\ (\bibinfo  {publisher} {Morgan Kaufmann Publishers Inc.},\
  \bibinfo {address} {San Francisco, CA, USA},\ \bibinfo {year} {1993})\ pp.\
  \bibinfo {pages} {737--744}\BibitemShut {NoStop}%
\bibitem [{\citenamefont {Adamson}\ \emph {et~al.}(2016)\citenamefont {Adamson}
  \emph {et~al.}}]{numi}%
  \BibitemOpen
  \bibfield  {author} {\bibinfo {author} {\bibfnamefont {P.}~\bibnamefont
  {Adamson}} \emph {et~al.},\ }\bibfield  {title} {\bibinfo {title} {{The NuMI
  Neutrino Beam}},\ }\href {https://doi.org/10.1016/j.nima.2015.08.063}
  {\bibfield  {journal} {\bibinfo  {journal} {Nucl. Instrum. Meth.}\ }\textbf
  {\bibinfo {volume} {A806}},\ \bibinfo {pages} {279} (\bibinfo {year}
  {2016})},\ \Eprint {https://arxiv.org/abs/1507.06690} {arXiv:1507.06690
  [physics.acc-ph]} \BibitemShut {NoStop}%
\bibitem [{\citenamefont {Pontecorvo}(1968)}]{PMNS}%
  \BibitemOpen
  \bibfield  {author} {\bibinfo {author} {\bibfnamefont {B.}~\bibnamefont
  {Pontecorvo}},\ }\bibfield  {title} {\bibinfo {title} {{Neutrino Experiments
  and the Problem of Conservation of Leptonic Charge}},\ }\href@noop {}
  {\bibfield  {journal} {\bibinfo  {journal} {Sov. Phys. JETP}\ }\textbf
  {\bibinfo {volume} {26}},\ \bibinfo {pages} {984} (\bibinfo {year} {1968})},\
  \bibinfo {note} {[Zh. Eksp. Teor. Fiz.53,1717(1967)]}\BibitemShut {NoStop}%
\bibitem [{\citenamefont {Ayres}\ \emph {et~al.}(2007)\citenamefont {Ayres}
  \emph {et~al.}}]{novadet}%
  \BibitemOpen
  \bibfield  {author} {\bibinfo {author} {\bibfnamefont {D.~S.}\ \bibnamefont
  {Ayres}} \emph {et~al.} (\bibinfo {collaboration} {NOvA}),\ }\bibfield
  {title} {\bibinfo {title} {{The NOvA Technical Design Report}}\ }\href
  {https://doi.org/10.2172/935497} {10.2172/935497} (\bibinfo {year}
  {2007})\BibitemShut {NoStop}%
\bibitem [{\citenamefont {Bichsel}\ \emph {et~al.}(2004)\citenamefont
  {Bichsel}, \citenamefont {Groom},\ and\ \citenamefont {Klein}}]{PDGpassage}%
  \BibitemOpen
  \bibfield  {author} {\bibinfo {author} {\bibfnamefont {H.}~\bibnamefont
  {Bichsel}}, \bibinfo {author} {\bibfnamefont {D.~E.}\ \bibnamefont {Groom}},
  and\ \bibinfo {author} {\bibfnamefont {S.~R.}\ \bibnamefont {Klein}},\
  }\bibfield  {title} {\bibinfo {title} {{Passage of particles through
  matter}},\ }\href@noop {} {\  (\bibinfo {year} {2004})}\BibitemShut {NoStop}%
\bibitem [{\citenamefont {Baird}\ \emph {et~al.}(2015)\citenamefont {Baird},
  \citenamefont {Bian}, \citenamefont {Messier}, \citenamefont {Niner},
  \citenamefont {Rocco},\ and\ \citenamefont {Sachdev}}]{novareco}%
  \BibitemOpen
  \bibfield  {author} {\bibinfo {author} {\bibfnamefont {M.}~\bibnamefont
  {Baird}}, \bibinfo {author} {\bibfnamefont {J.}~\bibnamefont {Bian}},
  \bibinfo {author} {\bibfnamefont {M.}~\bibnamefont {Messier}}, \bibinfo
  {author} {\bibfnamefont {E.}~\bibnamefont {Niner}}, \bibinfo {author}
  {\bibfnamefont {D.}~\bibnamefont {Rocco}}, and\ \bibinfo {author}
  {\bibfnamefont {K.}~\bibnamefont {Sachdev}},\ }\bibfield  {title} {\bibinfo
  {title} {{Event Reconstruction Techniques in NOvA}},\ }in\ \href
  {https://doi.org/10.1088/1742-6596/664/7/072035} {\emph {\bibinfo {booktitle}
  {{Proceedings, 21st International Conference on Computing in High Energy and
  Nuclear Physics (CHEP 2015): Okinawa, Japan, April 13-17, 2015}}}},\ Vol.\
  \bibinfo {volume} {664}\ (\bibinfo {year} {2015})\ p.\ \bibinfo {pages}
  {072035}\BibitemShut {NoStop}%
\bibitem [{\citenamefont {Dundar}\ and\ \citenamefont
  {Garcia{-}Dorado}(2017)}]{background}%
  \BibitemOpen
  \bibfield  {author} {\bibinfo {author} {\bibfnamefont {A.}~\bibnamefont
  {Dundar}}and\ \bibinfo {author} {\bibfnamefont {I.}~\bibnamefont
  {Garcia{-}Dorado}},\ }\bibfield  {title} {\bibinfo {title} {Context
  augmentation for convolutional neural networks},\ }\href
  {http://arxiv.org/abs/1712.01653} {\bibfield  {journal} {\bibinfo  {journal}
  {CoRR}\ }\textbf {\bibinfo {volume} {abs/1712.01653}} (\bibinfo {year}
  {2017})},\ \Eprint {https://arxiv.org/abs/1712.01653} {arXiv:1712.01653}
  \BibitemShut {NoStop}%
\bibitem [{\citenamefont {Tang}\ \emph {et~al.}(2015)\citenamefont {Tang},
  \citenamefont {Paluri}, \citenamefont {Fei{-}Fei}, \citenamefont {Fergus},\
  and\ \citenamefont {Bourdev}}]{imagecontext}%
  \BibitemOpen
  \bibfield  {author} {\bibinfo {author} {\bibfnamefont {K.~D.}\ \bibnamefont
  {Tang}}, \bibinfo {author} {\bibfnamefont {M.}~\bibnamefont {Paluri}},
  \bibinfo {author} {\bibfnamefont {L.}~\bibnamefont {Fei{-}Fei}}, \bibinfo
  {author} {\bibfnamefont {R.}~\bibnamefont {Fergus}}, and\ \bibinfo {author}
  {\bibfnamefont {L.~D.}\ \bibnamefont {Bourdev}},\ }\bibfield  {title}
  {\bibinfo {title} {Improving image classification with location context},\
  }\href {http://arxiv.org/abs/1505.03873} {\bibfield  {journal} {\bibinfo
  {journal} {CoRR}\ }\textbf {\bibinfo {volume} {abs/1505.03873}} (\bibinfo
  {year} {2015})},\ \Eprint {https://arxiv.org/abs/1505.03873}
  {arXiv:1505.03873} \BibitemShut {NoStop}%
\bibitem [{\citenamefont {Andreopoulos}\ \emph {et~al.}(2010)\citenamefont
  {Andreopoulos} \emph {et~al.}}]{genie}%
  \BibitemOpen
  \bibfield  {author} {\bibinfo {author} {\bibfnamefont {C.}~\bibnamefont
  {Andreopoulos}} \emph {et~al.},\ }\bibfield  {title} {\bibinfo {title} {{The
  GENIE Neutrino Monte Carlo Generator}},\ }\href
  {https://doi.org/10.1016/j.nima.2009.12.009} {\bibfield  {journal} {\bibinfo
  {journal} {Nucl. Instrum. Meth.}\ }\textbf {\bibinfo {volume} {A614}},\
  \bibinfo {pages} {87} (\bibinfo {year} {2010})},\ \Eprint
  {https://arxiv.org/abs/0905.2517} {arXiv:0905.2517 [hep-ph]} \BibitemShut
  {NoStop}%
\bibitem [{\citenamefont {Agostinelli}\ \emph {et~al.}(2003)\citenamefont
  {Agostinelli} \emph {et~al.}}]{geant}%
  \BibitemOpen
  \bibfield  {author} {\bibinfo {author} {\bibfnamefont {S.}~\bibnamefont
  {Agostinelli}} \emph {et~al.} (\bibinfo {collaboration} {GEANT4}),\
  }\bibfield  {title} {\bibinfo {title} {{GEANT4: A Simulation toolkit}},\
  }\href {https://doi.org/10.1016/S0168-9002(03)01368-8} {\bibfield  {journal}
  {\bibinfo  {journal} {Nucl. Instrum. Meth.}\ }\textbf {\bibinfo {volume}
  {A506}},\ \bibinfo {pages} {250} (\bibinfo {year} {2003})}\BibitemShut
  {NoStop}%
\bibitem [{\citenamefont {Aurisano}\ \emph {et~al.}(2015)\citenamefont
  {Aurisano}, \citenamefont {Backhouse}, \citenamefont {Hatcher}, \citenamefont
  {Mayer}, \citenamefont {Musser}, \citenamefont {Patterson}, \citenamefont
  {Schroeter},\ and\ \citenamefont {Sousa}}]{novasim}%
  \BibitemOpen
  \bibfield  {author} {\bibinfo {author} {\bibfnamefont {A.}~\bibnamefont
  {Aurisano}}, \bibinfo {author} {\bibfnamefont {C.}~\bibnamefont {Backhouse}},
  \bibinfo {author} {\bibfnamefont {R.}~\bibnamefont {Hatcher}}, \bibinfo
  {author} {\bibfnamefont {N.}~\bibnamefont {Mayer}}, \bibinfo {author}
  {\bibfnamefont {J.}~\bibnamefont {Musser}}, \bibinfo {author} {\bibfnamefont
  {R.}~\bibnamefont {Patterson}}, \bibinfo {author} {\bibfnamefont
  {R.}~\bibnamefont {Schroeter}}, and\ \bibinfo {author} {\bibfnamefont
  {A.}~\bibnamefont {Sousa}} (\bibinfo {collaboration} {NOvA}),\ }\bibfield
  {title} {\bibinfo {title} {{The NOvA simulation chain}},\ }\bibfield
  {booktitle} {\emph {\bibinfo {booktitle} {{Proceedings, 21st International
  Conference on Computing in High Energy and Nuclear Physics (CHEP 2015):
  Okinawa, Japan, April 13-17, 2015}}},\ }\href
  {https://doi.org/10.1088/1742-6596/664/7/072002} {\bibfield  {journal}
  {\bibinfo  {journal} {J. Phys. Conf. Ser.}\ }\textbf {\bibinfo {volume}
  {664}},\ \bibinfo {pages} {072002} (\bibinfo {year} {2015})}\BibitemShut
  {NoStop}%
\bibitem [{\citenamefont {Dunn}(1974)}]{fuzzyk}%
  \BibitemOpen
  \bibfield  {author} {\bibinfo {author} {\bibfnamefont {J.}~\bibnamefont
  {Dunn}},\ }\bibfield  {title} {\bibinfo {title} {A fuzzy relative of the
  isodata process and its use in detecting compact, well-separated clusters},\
  }\href@noop {} {\bibfield  {journal} {\bibinfo  {journal} {J. Cybern.}\
  }\textbf {\bibinfo {volume} {3}},\ \bibinfo {pages} {32} (\bibinfo {year}
  {1974})}\BibitemShut {NoStop}%
\bibitem [{\citenamefont {Niner}(2015)}]{ninerthesis}%
  \BibitemOpen
  \bibfield  {author} {\bibinfo {author} {\bibfnamefont {E.~D.}\ \bibnamefont
  {Niner}},\ }\emph {\bibinfo {title} {{Observation of Electron Neutrino
  Appearance in the NuMI Beam with the NOvA Experiment}}},\ \href
  {https://doi.org/10.2172/1221353} {Ph.D. thesis},\ \bibinfo  {school}
  {Indiana U.} (\bibinfo {year} {2015})\BibitemShut {NoStop}%
\bibitem [{\citenamefont {Szegedy}\ \emph {et~al.}(2014)\citenamefont
  {Szegedy}, \citenamefont {Liu}, \citenamefont {Jia}, \citenamefont
  {Sermanet}, \citenamefont {Reed}, \citenamefont {Anguelov}, \citenamefont
  {Erhan}, \citenamefont {Vanhoucke},\ and\ \citenamefont
  {Rabinovich}}]{googlenet}%
  \BibitemOpen
  \bibfield  {author} {\bibinfo {author} {\bibfnamefont {C.}~\bibnamefont
  {Szegedy}}, \bibinfo {author} {\bibfnamefont {W.}~\bibnamefont {Liu}},
  \bibinfo {author} {\bibfnamefont {Y.}~\bibnamefont {Jia}}, \bibinfo {author}
  {\bibfnamefont {P.}~\bibnamefont {Sermanet}}, \bibinfo {author}
  {\bibfnamefont {S.}~\bibnamefont {Reed}}, \bibinfo {author} {\bibfnamefont
  {D.}~\bibnamefont {Anguelov}}, \bibinfo {author} {\bibfnamefont
  {D.}~\bibnamefont {Erhan}}, \bibinfo {author} {\bibfnamefont
  {V.}~\bibnamefont {Vanhoucke}}, and\ \bibinfo {author} {\bibfnamefont
  {A.}~\bibnamefont {Rabinovich}},\ }\bibfield  {title} {\bibinfo {title}
  {{Going Deeper with Convolutions}},\ }\href@noop {} {\  (\bibinfo {year}
  {2014})},\ \Eprint {https://arxiv.org/abs/1409.4842} {arXiv:1409.4842
  [cs.CV]} \BibitemShut {NoStop}%
\bibitem [{\citenamefont {Psihas}(2018)}]{psihasthesis}%
  \BibitemOpen
  \bibfield  {author} {\bibinfo {author} {\bibfnamefont {F.}~\bibnamefont
  {Psihas}},\ }\emph {\bibinfo {title} {{Measurement of Long Baseline Neutrino
  Oscillations and Improvements from Deep Learning}}},\ \href
  {https://doi.org/10.2172/1437288} {Ph.D. thesis},\ \bibinfo  {school}
  {Indiana U.} (\bibinfo {year} {2018})\BibitemShut {NoStop}%
\bibitem [{\citenamefont {Jia}\ \emph {et~al.}(2014)\citenamefont {Jia},
  \citenamefont {Shelhamer}, \citenamefont {Donahue}, \citenamefont {Karayev},
  \citenamefont {Long}, \citenamefont {Girshick}, \citenamefont {Guadarrama},\
  and\ \citenamefont {Darrell}}]{caffe}%
  \BibitemOpen
  \bibfield  {author} {\bibinfo {author} {\bibfnamefont {Y.}~\bibnamefont
  {Jia}}, \bibinfo {author} {\bibfnamefont {E.}~\bibnamefont {Shelhamer}},
  \bibinfo {author} {\bibfnamefont {J.}~\bibnamefont {Donahue}}, \bibinfo
  {author} {\bibfnamefont {S.}~\bibnamefont {Karayev}}, \bibinfo {author}
  {\bibfnamefont {J.}~\bibnamefont {Long}}, \bibinfo {author} {\bibfnamefont
  {R.}~\bibnamefont {Girshick}}, \bibinfo {author} {\bibfnamefont
  {S.}~\bibnamefont {Guadarrama}}, and\ \bibinfo {author} {\bibfnamefont
  {T.}~\bibnamefont {Darrell}},\ }\bibfield  {title} {\bibinfo {title} {Caffe:
  Convolutional architecture for fast feature embedding},\ }\href@noop {}
  {\bibfield  {journal} {\bibinfo  {journal} {arXiv preprint arXiv:1408.5093}\
  } (\bibinfo {year} {2014})}\BibitemShut {NoStop}%
\bibitem [{\citenamefont {Brun}\ and\ \citenamefont {Rademakers}(1996)}]{ROOT}%
  \BibitemOpen
  \bibfield  {author} {\bibinfo {author} {\bibfnamefont {R.}~\bibnamefont
  {Brun}}and\ \bibinfo {author} {\bibfnamefont {F.}~\bibnamefont
  {Rademakers}},\ }\bibfield  {title} {\bibinfo {title} {Root - an object
  oriented data analysis framework},\ }in\ \href@noop {} {\emph {\bibinfo
  {booktitle} {AIHENP'96 Workshop, Lausane}}},\ Vol.\ \bibinfo {volume} {389}\
  (\bibinfo {year} {1996})\ pp.\ \bibinfo {pages} {81--86}\BibitemShut
  {NoStop}%
\bibitem [{\citenamefont {Green}\ \emph {et~al.}(2012)\citenamefont {Green},
  \citenamefont {Kowalkowski}, \citenamefont {Paterno}, \citenamefont
  {Fischler}, \citenamefont {Garren},\ and\ \citenamefont {Lu}}]{ART}%
  \BibitemOpen
  \bibfield  {author} {\bibinfo {author} {\bibfnamefont {C.}~\bibnamefont
  {Green}}, \bibinfo {author} {\bibfnamefont {J.}~\bibnamefont {Kowalkowski}},
  \bibinfo {author} {\bibfnamefont {M.}~\bibnamefont {Paterno}}, \bibinfo
  {author} {\bibfnamefont {M.}~\bibnamefont {Fischler}}, \bibinfo {author}
  {\bibfnamefont {L.}~\bibnamefont {Garren}}, and\ \bibinfo {author}
  {\bibfnamefont {Q.}~\bibnamefont {Lu}},\ }\bibfield  {title} {\bibinfo
  {title} {{The Art Framework}},\ }\bibfield  {booktitle} {\emph {\bibinfo
  {booktitle} {{Proceedings, 19th International Conference on Computing in High
  Energy and Nuclear Physics (CHEP 2012): New York, USA, May 21-25, 2012}}},\
  }\href {https://doi.org/10.1088/1742-6596/396/2/022020} {\bibfield  {journal}
  {\bibinfo  {journal} {J. Phys. Conf. Ser.}\ }\textbf {\bibinfo {volume}
  {396}},\ \bibinfo {pages} {022020} (\bibinfo {year} {2012})}\BibitemShut
  {NoStop}%
\bibitem [{\citenamefont {van~der Maaten}\ and\ \citenamefont
  {Hinton}(2008)}]{tsne}%
  \BibitemOpen
  \bibfield  {author} {\bibinfo {author} {\bibfnamefont {L.}~\bibnamefont
  {van~der Maaten}}and\ \bibinfo {author} {\bibfnamefont {G.}~\bibnamefont
  {Hinton}},\ }\bibfield  {title} {\bibinfo {title} {Visualizing
  high-dimensional data using t-sne},\ }\href@noop {} {\  (\bibinfo {year}
  {2008})}\BibitemShut {NoStop}%
\bibitem [{\citenamefont {MUETHER}(2018)}]{neutrinoNuNC}%
  \BibitemOpen
  \bibfield  {author} {\bibinfo {author} {\bibfnamefont {M.}~\bibnamefont
  {MUETHER}},\ }\href {https://doi.org/10.5281/zenodo.1300753} {\bibinfo
  {title} {{Status of the Neutrino-Induced Neutral Current Neutral Pion
  Production Cross Section Measurement from NOvA}}} (\bibinfo {year}
  {2018})\BibitemShut {NoStop}%
\bibitem [{\citenamefont {BAIRD}\ and\ \citenamefont
  {SMITH}(2018)}]{neutrinoEnergy}%
  \BibitemOpen
  \bibfield  {author} {\bibinfo {author} {\bibfnamefont {M.}~\bibnamefont
  {BAIRD}}and\ \bibinfo {author} {\bibfnamefont {E.}~\bibnamefont {SMITH}},\
  }\href {https://doi.org/10.5281/zenodo.1301120} {\bibinfo {title}
  {{Reconstructing Neutrino Energies with the NOvA Detectors}}} (\bibinfo
  {year} {2018})\BibitemShut {NoStop}%
\bibitem [{\citenamefont {LANG}(2018)}]{testbeam}%
  \BibitemOpen
  \bibfield  {author} {\bibinfo {author} {\bibfnamefont {K.}~\bibnamefont
  {LANG}},\ }\href {https://doi.org/10.5281/zenodo.1300576} {\bibinfo {title}
  {The nova test beam program}} (\bibinfo {year} {2018})\BibitemShut {NoStop}%
\end{thebibliography}%

\end{document}